\newcommand{\meq}{\stackrel{?}{=}}
\newcommand{\low}[1]{{\stackrel{}{#1}}}
\newcommand{\sub}[1]{_{\stackrel{}{#1}}}
\newcommand{\rhog}{\rho\sub\gamma}
\newcommand{\lp}{{\stackrel{\leftarrow}{\partial}}}
\newcommand{\rp}{{\stackrel{\rightarrow}{\partial}}}
\newcommand{\lpq}{\stackrel{\leftarrow}{\partial}_{q}}

\newcommand{\rpq}{\stackrel{\rightarrow}{\partial}_{q}}

\newcommand{\lpp}{\stackrel{\leftarrow}{\partial}_{p}}
\newcommand{\rpp}{\stackrel{\rightarrow}{\partial}_{p}}

\newcommand{\pq}{{\partial}_{q}}
\newcommand{\pp}{{\partial}_{p}}
\newcommand{\ben}{\begin{equation}}
\newcommand{\een}{\end{equation}}
\newcommand{\bea}{\begin{eqnarray}}
\newcommand{\eea}{\end{eqnarray}}
\newcommand{\im}{{\,\rm Im}}
\newcommand{\nn}{\nonumber\\ }

\newcommand{\qq}{\qquad\qquad}
\newcommand{\QQ}{\qquad\qquad\qquad\qquad}

\newcommand{\infinity}{{\infty}} \newcommand{\w}{{\wedge}}
\newcommand{\cW}{{\cal W}} \newcommand{\cC}{{\cal C}}
\newcommand{\cE}{{\cal E}} \newcommand{\cO}{{\cal O}}
\newcommand{{\ga}}{{\gamma}} \newcommand{\stg}{{*_\gamma}}
\newcommand{\E}{{\rm Exp}} \newcommand{\stmg}{*_{-\gamma}}
\newcommand{\Lg}{{\cal L}_\gamma}
\newcommand{\tT}{{\tilde T}}
\newcommand{\tst}{{\tilde *}}

\input amssym.def

\def\R{{\Bbb R}}

\documentclass{article}
\usepackage{graphicx}
\usepackage{epsfig}
\begin{document}

\parskip=4pt \baselineskip=14pt %%%%%%%%%%%% Title %%%%%%%%%%

\title{\vskip-1cm On Wigner functions and a damped star product in dissipative phase-space quantum mechanics}
\author{B. Belchev, M.A. Walton\\\\ {\it Department of Physics, University of Lethbridge}\\ {\em Lethbridge, Alberta,
Canada\ \  T1K 3M4}\\\\ {\small borislav.belchev@uleth.ca, walton@uleth.ca}\\\\ }

\maketitle
%%%%%%%%%%%%%%%%%%%%%%%%%%%%%%%%%%%%%%%%%%%%%%%%%%%%%%%%%%%%%%%%%%%%%%%%%%%%%%%%%%%%%%%%%%%%%%%%%%%%%%%%%%%%%%%%
\begin{abstract}

Dito and Turrubiates recently introduced an interesting model of the dissipative quantum mechanics of a damped harmonic
oscillator in phase space. Its key ingredient is a non-Hermitian deformation of the Moyal star product with the damping
constant as deformation parameter. We compare the Dito-Turrubiates scheme with phase-space quantum mechanics (or deformation quantization) 
based on other star products,
and extend it to incorporate Wigner functions. The 
deformed (or damped) star product is related to a complex Hamiltonian, and so necessitates a modified equation of motion involving complex
conjugation. We find that with this change the Wigner function satisfies the classical equation of motion. This seems
appropriate since non-dissipative systems with quadratic Hamiltonians share this property.

\end{abstract}

\vfill\eject
%%%%%%%%%%%%%%%%%%%%%%%%%%%%%%%%%%%%%%%%%%%%%%%%%%%%%%%%%%%%%%%%%%%%%%%%%%%%%%%%%%%%%%%%%%%%%%%%%%%%%%%%%%%%%%%%
\section{ Introduction } 

The quantum mechanics of dissipative systems has been studied intensively; for reviews, see \cite{D,Raz,W}. Work started soon after the birth of quantum mechanics and continues today.

Perhaps the most fundamental approach is to consider the system of interest as interacting with an appropriate reservoir. Then the well-known quantization methods that are valid for non-dissipative, closed systems can be applied to the system plus reservoir as a whole. Effective equations of motion for the system can then be found by integrating  out the reservoir degrees of freedom, while making appropriate physical assumptions.

Another technique is to work backwards, and derive the effective equations by adapting quantization procedures to non-dissipative systems.  The adapted quantization procedures should, in the end, agree with more fundamental treatments. Provided they do, they would be helpful, as shortcuts for the more fundamental derivations, and possibly more.

Canonical (operator) quantization has been adapted to dissipative systems either by using time-dependent Hamiltonians, complex Hamiltonians, or by modifying the canonical Poisson brackets, and the corresponding operator commutators.\footnote{\, Some relatively recent works with the latter approach are \cite{K,G,E,BEMS}.} 

We concern ourselves here, however, with a different quantization method: quantum mechanics in phase space, or deformation quantization (see Sect. 2). Specifically, we study Dito and Turrubiates \cite{DT} recent adaptation of deformation quantization to the paradigmatic dissipative system, the damped harmonic oscillator. 

The important innovation they introduce is a non-Hermitian $\gamma$-deformation of the Moyal star product of non-dissipative deformation quantization, where $\gamma$ denotes the damping constant. Their damped star product is built in turn on a $\gamma$-deformation of the classical Poisson bracket, and so recalls the modified brackets of canonical operator quantum mechanics adapted to dissipation that were just mentioned. Perhaps this is not surprising, since canonical classical mechanics is a key ingredient of deformation quantization;  the algebra of quantum observables is described as an $\hbar$-deformation of the classical Poisson algebra on phase space. Unlike in other schemes, the one proposed by Dito and Turrubiates {\it only} modifies the classical bracket and thereby the corresponding quantum star product -- it uses the undamped Hamiltonian, for example. The Dito-Turribiates proposal is at least economical, since in other formulations involving modified brackets, extra structure must be input.

This paper is organized as follows. The next section is a brief review of ordinary (undamped) quantum mechanics in phase space. Section 3 studies the Dito-Turrubiates scheme and extends it to include Wigner functions and their evolution. The final section is a discussion, consisting of a concise summary, and a list of some of the many questions that remain.

\section{ Phase-space quantum mechanics }

We now give a brief review of phase-space quantum mechanics, to establish our notation and provide the essential
background for next section's discussion of the Dito-Turrubiates scheme. Throughout this paper our treatment will be
limited to one non-relativistic particle travelling on the real line with coordinate $q$, and conjugate momentum $p$.

For pedagogical introductions to phase-space quantum mechanics (or deformation quantization), see \cite{HWW,HH}, for
examples; and for more advanced reviews, see the book \cite{ZFC}, and references therein, including the pioneering work  \cite{BFFLS}.

In our context, operators will be expressible entirely in terms of the operator position $\hat q$ and momentum $\hat
p$. These more general operators will also be indicated by carets. Phase-space quantum mechanics has a direct
relationship with the more commonly used operator quantum mechanics. The Wigner transform $\cW$ maps operators to
functions and distributions on phase space: \ben \cW\left(\, \hat f \,\right)\ =\ f(p,q)\ .\label{Wigtr}\een Here $\hat
f$ is an operator, and $f(p,q)$ is the corresponding phase-space distribution, sometimes called the {\it symbol} of
$\hat f$. The most important property of the Wigner transform is encoded in\ben \cW(\, \hat f\, \hat g \,)\ =\ (\cW
\hat f)\,*\, (\cW\hat g)\ . \label{Whom}\een Consequently, operators can be Wigner-mapped to symbols whose algebra is
homomorphic to the original operator algebra, as long as the symbols are multiplied using the so-called star-product
($*$-product) \ben *\ =\ \exp\Big[\, \frac{i\hbar}2 (\lpq \rpp - \lpp \rpq ) \,\Big]\ .\label{Moystar}\een Operators
can therefore be replaced by functions/distributions on phase space, and quantum mechanics can be done in phase space,
without reference to operators.

The Moyal star product $*$ inherits associativity from the operator product, by (\ref{Whom}). Since $(\hat f\hat g)^\dagger = \hat g^\dagger \hat f^\dagger$, we also have \ben \overline{\left(\, f\, *\, g\,\right)}\ =\ \overline{g}\, *\, \overline{f}\ .    \label{MoyHerm}\een Here $\overline{f}$ indicates the complex conjugate of $f$. We say that $*$ is Hermitian.

The inverse of the Wigner transform is familiar in operator quantization. The so-called Weyl map $\cW^{-1}$, satisfies
\ben \cW^{-1}(\, f\, *\, g \,)\ =\ (\cW^{-1} f)\, (\cW^{-1} g)\ . \label{Winvhom}\een The image of a polynomial in $q$
and $p$ of this Weyl map is the corresponding Weyl-ordered operator.  The famous Moyal star product $*$ of (\ref{Moystar}) and (\ref{Winvhom}) is therefore intimately related to Weyl ordering. With other operator orderings, other $*$-products arise.

The operator encoding information about the quantum state is the density matrix, and
its Wigner transform is the central object of phase-space quantum mechanics. Up to normalization, it is the celebrated
Wigner function.

To consider evolution, start with the Liouville Theorem for the classical phase-space density $\rho_c$: \ben
\frac{d\rho_c}{dt}\ =\ \frac{\partial \rho_c}{\partial t}\ +\ \{ \rho_c, H \}\ =\ 0\ . \label{Liouvillerhc}\een The
Dirac correspondence then yields the equation of motion of the density operator (matrix) $\hat \rho$ \ben  i\hbar\,
\frac{\partial\hat\rho}{\partial t}\ +\ [\, \hat\rho, \hat H \,]\  =\ 0\ . \label{evhr}\een The Wigner transform of
this is the evolution equation for the Wigner function $\rho=\rho(p,q;t)$ \ben i\hbar\, \frac{\partial\rho}{\partial
t}\ +\ [\,  \rho\, ,\, H \,]_*\ =\ 0\ , \label{evWig}\een where the $*$-commutator is \ben [\, a\, ,\, b\,]_*\ =\
a\,*\, b\ -\ b\,*\, a\ . \label{stcomm}\een Equivalently, the phase-space version of the Dirac quantization rule, \ben
\{a, b\}\ \rightarrow\ \frac{[\,a\, ,\,b\,]_*}{i\hbar}\ , \label{Dirps}\een leads directly from (\ref{Liouvillerhc}) to
(\ref{evWig}).

Defining \ben {\rm ad}\sub f[*]\,g\ :=\ [\,f\, ,\,g\,]_*\ , \label{defadst}\een (\ref{evWig}) can be rewritten as \ben
i\hbar\, \frac{\partial\rho}{\partial t}\ =\ {\rm ad}\sub H[*]\,\rho \ =:\ i\hbar{\cal L}\, \rho\  .
\label{evWigad}\een The solution is just \ben \rho(p,q;t)\ =\ \exp\left\{\,\frac{-it}{\hbar}\, {\rm ad}\sub H[*]
\,\right\}\, \rho(p,q;0)\ =\ \exp\left\{\,t\,{\cal L} \,\right\}\, \rho(p,q;0)\ . \label{expadst}\een Using the
associativity of $*$-multiplication, this  reduces to \ben \rho(p,q;t)\ =\ U(p,q;t)\, *\, \rho(p,q;0)\, *\, U(p,q;-t)\
, \label{UrhoU}\een with \ben U(p,q;t)\ =\ \sum_{n=0}^\infinity\, \frac1{n!}\,\left(\frac{-itH}{\hbar}\right)^{*n}\ =\
\E[*]\left(\frac{-itH}{\hbar}\right)\ .\label{Utx}\een  Here \ben \E[*](a)\ :=\ \sum_{n=0}^\infinity \frac 1{n!}\,
a^{*n}
  \label{stExp}\een denotes the so-called $*$-exponential.
Note that $\overline{U(p,q;t)} = U(p,q;-t)$, and \ben \overline{U(p,q;t)}*U(p,q;t)\ =\ U(p,q;t)*\overline{U(p,q;t)}\ =\
1\ .\label{bUstU}\een

Alternatively, substituting (\ref{UrhoU}) into (\ref{evWig}) yields the dynamical equation \ben i\hbar\partial_t\,
U(p,q;t)\ =\ H\,*\, U(p,q;t)\ , \label{dynU}\een and its complex conjugate. Its solution, the symbol of the propagator,
is (\ref{Utx}). The spectrum of energies and the Wigner functions of stationary states can then be found by the
spectral decomposition (or Fourier-Dirichlet expansion) \ben U(p,q;t)\ =\ \sum_E\, \rho\sub E(p,q)\, e^{-iEt/\hbar}\ ,
\label{Ustat}\een where \ben  H\, *\, \rho\sub E\ =\ \rho\sub E\, *\, H\ =\ E\, \rho\sub E\ . \label{HrrHEr}\een

For
the simple harmonic oscillator, with Hamiltonian \ben H\ =\ \frac{p^2}{2m}\ +\ \frac12 m\omega^2 q^2\ ,
\label{dhoH}\een the well-known results are  \ben  U(p,q;t)\ =\ \frac{1}{\cos(\omega t/2)}\, \exp\left\{\,
\frac{2H\tan(\omega t/2)}{i\hbar\omega} \,\right\}\ ,\label{Usho}\een and \ben \rho\sub{E_n}\ =\ 2(-1)^n\,L_n\left(
\frac{4H}{\hbar\omega} \right)\, \exp\left( -\frac{2H}{\hbar\omega} \right)\ ,\label{rEsho}\een where $E_n=
\hbar\omega(n + 1/2)$, and $L_n$ is the $n$-th Laguerre polynomial.

The time evolution invokes more than the pure-state, or diagonal Wigner functions, however. We need the off-diagonal
\ben \rho_\low{E,E'}\ =\ \frac1{2\pi\hbar}\, \cW\left(\, |E\rangle\langle E'| \,\right)\ ,\label{offEEp}\een satisfying
the $*$-eigen equations \ben H\,*\,\rho\sub{E,E'}\ =\ E\, \rho\sub{E,E'}\ ,\ \ \ \rho\sub{E,E'}\, *\, H\ =\
E'\,\rho\sub{E,E'}\ ;\label{rhoEEp}\een so that $\rho\sub{E,E} = \rho\sub E$. Assuming that the $*$-eigen functions
(\ref{rhoEEp}) are complete, we expand \ben  \rho(p,q;t)\ =\ \sum_{E,E'}\, R\sub{E,E'}(t)\, \rho\sub{E,E'}(p,q)\ .
 \label{Wig0tt}\een Substituting into the equation of motion then gives
the time-evolved Wigner function \ben \rho(p,q;t)\ =\ \sum_{E,E'}\, R\sub{E,E'}(0)\, \exp\left[\frac{-i(E-E')t}{\hbar}
\right]\, \rho\sub{E,E'}(p,q)\ .
 \label{Wig0t}\een

To close this section, we will discuss alternatives to the Moyal star product (see \cite{HH} and \cite{BFFLS}, e.g.). First, consider operator-ordering ambiguities. Since the Moyal product is intimately related to Weyl ordering, a different operator ordering will lead to a different star product. For example, if we use the so-called standard ordering, where every $\hat q$ is placed to the left of every $\hat p$, we find instead the standard star product \ben *\sub{S}\ =\ e^{i\hbar\lpq\rpp}\ =\ *\,\, e^{\frac{i\hbar}2\,(\lpq\rpp+\lpp\rpq)}\ .\label{standardstar}\een Since the two orderings can be simply related, there is a nice relation between the two star products: \ben T\sub{S}(\, f\,*\sub{S}\, g \,)\ =\ (T\sub{S}f)\, *\, (T\sub{S}g)\ , \label{Rstars}\een involving the invertible {\it transition operator} \ben T\sub{S}\ =\
\exp\left\{\, -\frac{i\hbar}2\,\pq\,\pp \,\right\}\ . \label{Rform}\een An ordering change preserves a bi-grading in the powers of $p$ and
$q$. By the Heisenberg commutation relation, $\hbar$ has bi-grade (1,1) and so that of $T\sub{S}$ is (0,0).

This kind of connection between different star products can be extended. In general, a star product $\tst$  is considered equivalent to the Moyal one $*$, if we have \ben \tT(\, f\,*\, g \,)\ =\
(\tT f)\, \tilde *\, (\tT g)\ .\label{Qstarp}\een Here $\tT$ stands for an invertible transition operator. If $\tilde *$, like $*$, is an $\hbar$-deformation of the point-wise product of functions, then \ben \tT\ =\
1\ +\ \cO(\hbar)\ .\label{Qoneh}\een Associativity of $\tilde *$ is guaranteed by the relation (\ref{Qstarp}), if $\tT$ is invertible: \ben
f\,\tilde *\, g\,\tilde * h\ =\ \tT\Big(\, (\tT^{-1}f)\,*\, (\tT^{-1}g)\, *\, (\tT^{-1}h) \,\Big)\ . \label{assoctstar}\een If $\tT$ is real, then $\tilde *$ will be Hermitian, as $*$ is. $*\sub{S}$ is an example of a non-Hermitian product, since $T\sub{S}$ is not real.

By the Liebniz product rule, the equivalence relation (\ref{Qstarp}) can be rewritten as \ben \tilde *\ =\ *\ \tT^{-1}[\lp]\ \tT[\lp+\rp]\  \tT^{-1}[\rp]\  .\label{tststT}\een Here  $\partial$ stands for either $\partial_q$ or $\partial_p$,  \ben \tT[\lp+\rp]\ :=  \tT\,\vert_{\partial\to \lp+\rp}\ \ , \label{defTpart}\een and similarly for $\tT^{-1}[\lp]$ and $\tT^{-1}[\rp]$. For example, \ben T\sub{S}[\lp+\rp]\ =\
\exp\left\{\, -\frac{i\hbar}2\,(\lpq+\rpq)\,(\lpp+\rpp) \,\right\}\ , \label{RformT}\een by (\ref{Rform}).

Eqn. (\ref{tststT}) makes it clear that equivalent star products $*$ and $\tilde *$ only differ by a factor that is left-right symmetric. Consequently, it can be shown that \ben \lim_{\hbar\to 0}\frac{[f,g]_{\tilde *}}{i\hbar}\ =\ \lim_{\hbar\to
0}\,\frac{[f,g]_{*}}{i\hbar}\ =\ \{f,g\}\ ,\label{stplimh}\een the Poisson bracket.

The equivalence between $\tilde *$ and $*$ defined by (\ref{Qstarp}) is known as c-equivalence. It is important because the Moyal $*$-product (\ref{Moystar}) is the
unique, associative deformation of the point-wise multiplication of functions on $\R^2$ (and $\R^{2n}$), up to  c-equivalence.

The c in c-equivalence stands for cohomological. As such, it is a mathematical, rather than a physical equivalence. For example, consider the Husimi star product $*\sub{H}$ (see \cite{T}, e.g.) \ben  *\sub H\ =\  *\,\exp\left\{\, \frac\hbar 2\,(s^2\,\lpq\rpq \,+\, \frac 1{s^2}\, \lpp\rpp) \,\right\}\  ,\label{starH}\een related to $*$ by the transition operator  \ben T\sub{H}\ =\ \exp\left\{\frac\hbar 4 \big( s^2\partial_q^{\, 2}+ \frac 1{s^2}
\partial_p^{\, 2} \big) \right\}\  .  \label{TsubH}\een It is also easy to see that $*\sub H$ satisfies the condition (\ref{stplimh}). But $*\sub H$ is used with the Husimi phase space distribution \ben
\rho\sub H(p,q;t)\  :=\ T\sub{H}\,\rho(p,q;t)\  \label{HusimiWT}\een which does not encode precisely the same physics as the Wigner function $\rho(p,q;t)$. Eqn. (\ref{HusimiWT}) can be rewritten as \bea \rho\sub H(p,q;t)\ =\ \frac
1 {\pi\hbar}\, \int dp'\, dq'\, \rho(p',q';t)\QQ\qq\nn \QQ \times\ \exp\left\{\, -\frac 1 \hbar\, \left[
\frac{(q-q')^2}{s^2}\, +\, s^2(p-p')^2 \right] \,\right\}\ , \label{smoothHW}\eea
indicating that the Husimi
distribution is a smoothed version of the Wigner function, coarse-grained by a squeezed\footnote{\, More precisely, the
distribution functions for general  $s$ were introduced in \cite{Ca,Ra}, while the Husimi distribution has $s=1$. We call the Gaussian weighting of (\ref{smoothHW}) squeezed because it is proportional to the Wigner
transform of a squeezed state.} Gaussian weighting in phase space. So, although $*\sub H$ is c-equivalent to the Moyal product, it describes different physics.

\section{ The damped harmonic oscillator in phase-space quantum mechanics }

We start in this section with a  review of the Dito and Turrubiates \cite{DT} model  of damping in a quantum
harmonic oscillator in phase space, with  comments added.  A new contribution will then be described: our
incorporation of Wigner functions and their evolution. First, a na\" ive proposal will be examined, and rejected, since
it leads to an unacceptable evolution equation. The equation of motion will then be modified, and explained. Finally,
its consequences are outlined. One such consequence is that the Wigner function of the damped harmonic oscillator
follows the canonical flow. This property is also common to all non-dissipative systems having quadratic Hamiltonians,
including the simple (undamped) harmonic oscillator. Since the damped and undamped harmonic oscillator and such systems
are treated similarly in some other approaches, this seems physically reasonable.

Dito and Turrubiates \cite{DT} describe the damped harmonic oscillator using the Hamiltonian of the undamped simple
harmonic oscillator. In a sense then, they describe its dissipation as kinematics. More precisely, the dissipation is
encoded in a deformation of the classical Poisson bracket of the harmonic oscillator and its consequent quantum
$*$-product.

The initial observation is that the classical equations of motion of the damped harmonic oscillator can be written
as \ben \dot q\ =\ \{q,H\}\sub\gamma\ =\ p/m\ ;\ \ \dot p\ =\ \{p,H\}\sub\gamma\ =\ -m\omega^2q\ -\ 2\gamma p\
,\label{dhoem}\een using the {\it undamped} Hamiltonian (\ref{dhoH}). The price paid is that we must use a deformed Poisson bracket \bea \{\, f,g \,\}\sub\gamma\ :=\ \{f,g\}\ -\ 2\gamma
m\,\partial_pf\,\partial_pg\ \ \qq\nn  =\ f\left(\,\lpq\,\rpp\ -\ \lpp\,\rpq\ -\ 2\gamma
m\,\lpp\,\rpp\,\right)g\ ,\label{Mbracket}\eea with the damping constant $\gamma$ as deformation parameter. Notice
that the deformed bracket is no longer skew,\footnote{\, The deformed bracket can be obtained from the Poisson bracket by the substitutions $\lpq\to\lpq-2\alpha m\gamma\lpp$ and $\rpq\to\rpq+2\beta m\gamma\rpp$, for any real $\alpha$ and $\beta$ such that $\alpha+\beta=1$. The sign difference in front of $\gamma$ between the substitutions for the left- and right-acting derivatives foreshadows our proposal (\ref{sarhoevnobar}).} and doesn't obey the Jacobi identity.

This result applies to a general class of classical systems:  if we use any Hamiltonian of the form \ben \tilde H\ =\  \frac{p^2}{2m}\ +\ V(q)\ ,\label{tHam}\een  with an arbitrary potential $V(q)$, the same damping term $2m\gamma \dot q$ is still the only modification of the equation of motion for $q(t)$. We will restrict attention here to the simple harmonic oscillator and its damped version, however.

Just as the Moyal product is related to the Poisson bracket, the Dito-Turrubiates damped $*$-product is obtained by
exponentiation of the deformed classical bracket (\ref{Mbracket}): \ben   \stg\ :=\ \exp\Big[\, \frac{i\hbar}2 (\lpq
\rpp - \lpp \rpq - 2\gamma m \lpp \rpp ) \,\Big]\ =\ *\, e^{-i\hbar\gamma m \lpp\rpp }\ .\label{DTstar}\een Why
exponentiate? In the undamped case, it is necessary for the homomorphism of the $*$-algebra with the operator algebra.
Also, the Moyal $*$-product is the unique associative deformation of the point-wise multiplication of functions on
$\R^{2n}$, up to isomorphism (by transition operators -- see below). No similar result for the damped product is known;
neither is the operator algebra for the damped case. Exponentiation is therefore an assumption.

On the other hand, suppose that  we posit \ben \stg\ :=\ F\left(\, \frac{i\hbar}2 (\lpq \rpp -
\lpp \rpq - 2\gamma m \lpp \rpp ) \,\right)\ ,\label{DTstarF}\een for some function $F$. Since $\{\cdot,\cdot\}_\gamma
\to \{\cdot,\cdot\}$ as $\gamma\to 0$, requiring that $\lim_{\gamma\to 0}\stg = *$ selects the exponential function.
The possibility that the form (\ref{DTstarF}) is too restrictive remains, however.

One key result of \cite{DT} is  \ben T(\, f\,*\, g\,)\ =\ (Tf)\,\stg\, (Tg)\ , \label{dstarT}\een with  \ben T\ =\
\exp\Big(\, \frac{-i\hbar m\gamma}2 \, \partial_p^{\,2} \,\Big)\ . \label{defT}\een   This can be proved easily using (\ref{tststT}). That is, the Dito-Turrubiates damped star product is c-equivalent to the Moyal star product. As  discussed in the previous section, c-equivalence does not imply physical equivalence, and so the damped star product has the potential to describe the dissipation of a damped harmonic oscillator.

So, what effect does the Dito-Turrubiates transition operator $T$ have?  First, it is clear that $T$ does more than change the ordering, since it has no fixed bi-grade in the powers of $p$ and $q$. Its effect is therefore more profound. An important example is   \ben T\left(\, \frac{p^2}{2m}\ +\
V(q) \,\right)\ =\ \frac{p^2}{2m}\ +\ V(q)\ -\ i\frac{\hbar\gamma}2\ ;\label{TKE}\een (see also (\ref{gUtxHc})\,).\footnote{\,  This
last result is also interesting in its own right, since it recalls a different approach to damped systems that uses
complex Hamiltonians (see \cite{D,Raz}, and also \cite{Raj}).} $T$  transforms a real Hamiltonian into a complex one.

For such a conversion to be possible, it is necessary, but not sufficient, that $\overline{T}\not= T$. Another consequence of a non-real transition operator is that $\stg$ is not Hermitian: \ben \overline{a \, \stg\, b}\ =\ \overline{b}\, *_{-\gamma}\, \overline{a}\ \not=\ \overline{b}\, *_{\gamma}\, \overline{a}\ .\label{dstarcc}\een

Its c-equivalence with the Moyal star product ensures that the damped star product $\stg$ is associative, for any value of the damping constant $\gamma$. For example, $\stmg$ will be useful later because of (\ref{dstarcc}), and it is also associative. However, problems exist if any two different damping parameters $\gamma$ and $\gamma'$ are used.  Care must be taken with regard to the order of multiplications. Clearly, \ben  f\,\stg\, \left(\,
1\,*_{\gamma'}\, g\,\right)\ \not=\ \left(\, f\, \stg\, 1\, \right)\, *_{\gamma'}\, g\ , \label{foneh}\een if
$\gamma\not= \gamma'$, for example. A unique product such as $a\stg b\stmg c$ does not (automatically) exist.

Use of a na\"ive substitution \ben
\{a, b\}_\gamma\ {\stackrel{?}{\rightarrow}}\ \frac{[\,a\, ,\,b\,]_\stg 
}{i\hbar}\  \label{Dirpsnaive}\een leads to \ben 0\ =\ \frac{\partial\rhog}{\partial t}\ +\ \frac
1{i\hbar} [ \rhog , H]_{\stg}\  , \qq \label{Lrhog}\een where $\rhog$ indicates the Wigner function of the {\it damped}
harmonic oscillator. This equation of motion integrates to \ben \rho_\gamma(p,q;t)\ =\ U_\gamma(p,q;t)\, \stg\,
\rho_\gamma(p,q;0)\, \stg\, U_\gamma(p,q;-t)\ . \label{gUrhoU}\een Here \ben U_\gamma(p,q;t)\ :=\
\sum_{n=0}^\infinity\, \frac1{n!}\,\left(\frac{-itH}{\hbar}\right)^{\stg n}\ =\
\E[\stg]\left(\frac{-itH}{\hbar}\right)\ \label{gUtx}\een satisfies the damped analogue of (\ref{dynU}), i.e. \ben
i\hbar\partial_t\, U_\gamma(p,q;t)\ =\ H\,\stg\, U_\gamma(p,q;t)\ . \label{dynUg}\een

Dito-Turrubiates solve (\ref{dynUg}), without discussing the evolution of Wigner functions. They then Fourier-Dirichlet
expand the solution to find the spectrum of eigenvalues and eigenfunctions of \ben H\,\stg\, \rho\sub{\gamma,E}\ =\
E\sub\gamma\,\rho\sub{\gamma,E}\ .\label{HstgrEr}\een Using (\ref{dstarT}), one finds \bea U_\gamma(p,q;t)\ =\
\E[\stg]\left(\frac{-itH}{\hbar}\right)\ =\ T\Big(\, \E[*]\left(\frac{-itT^{-1}(H)}{\hbar}\right) \,\Big)\nn \ =\
T\Big(\, \E[*]\left(\frac{-it[H+i\hbar\gamma/2] }{\hbar}\right) \,\Big)\ =\ e^{\gamma t/2}\, T\Big( U(p,q;t)\Big)\ .
\label{gUtxHc}\eea Using (\ref{Usho}) and (\ref{defT}), this gives \cite{DT} \bea  U\sub\gamma(p,q;t)\ =\
\frac{\exp(\gamma t/2)}{\cos(\omega t/2)\left[ 1+\frac{2\gamma}{\omega}\tan(\omega t/2) \right]}\,\qq\QQ\nn \times\,
\exp\left\{\, \frac{2\tan(\omega t/2)}{i\hbar\omega}\,\left( \frac{p^2}{2m[ 1+\frac{2\gamma}{\omega}\tan(\omega t/2)
]}\, +\, \frac 1 2 m\omega^2q^2 \right) \,\right\}\ .\label{Ugdho}\eea  The expansion of $U_\gamma$ gives the solutions
of (\ref{HstgrEr}) in terms of the simple harmonic oscillator analogues: \ben \rho\sub{\gamma, n}\ =\ T\left(
\rho\sub{E_n} \right)\ , \label{rgEnTrEn}\een with \ben  E\sub{\gamma,n}\ =\ E_n\ +\ i\frac{\hbar\gamma}2\ =\
\frac{\hbar}2\left[ (2n+1)\omega\ +\ i\gamma \right]\ . \label{Egn}\een For example, for $n=0$ Dito and Turrubiates
\cite{DT} find \bea \rho\sub{E\sub{\gamma,0}}\ =\ \frac{2}{\sqrt{1-2i\gamma/\omega}} \QQ\QQ\nn \times\, \exp\left\{\,
-\frac{2}{\hbar\omega}\,\left( \frac{p^2}{2m[ 1-2i\gamma/\omega ]}\, +\, \frac 1 2 m\omega^2q^2 \right) \,\right\}\
.\label{rgEdhoz}\eea

However, there is a fundamental problem with the basic equation of motion, eqn. (\ref{Lrhog}). For the damped harmonic
oscillator it yields \ben 0\ =\ \frac{\partial\rhog}{\partial t}\ +\ \frac 1{i\hbar} [ \rhog , H]_\stg\ =\
\frac{\partial\rhog}{\partial t}\ +\ \frac 1{i\hbar} [ \rhog , H]_*\ +\ i\gamma\hbar\,\pp\pq\rhog\ .\qq
\label{Lrhogdet}\een Notice that the term proportional to the damping constant $\gamma$ is not real. But the
eigenvalues of the density matrix are the populations, and they must be real. Time evolution must preserve the reality
of the density matrix. The imaginary damping term in (\ref{Lrhogdet}) means that a real density matrix does not remain
real as it evolves.

An equally important flaw is revealed by the limit \ben \lim_{\hbar\rightarrow 0}\, \frac 1{i\hbar} [ H , \rhog]_\stg\
=\ \{ H, \rhog \}\ \not=\ \{ H, \rhog \}_\gamma\ .\label{limhstg}\een This is a direct consequence of the c-equivalence of $\stg$ and $*$. But this shows that there is no damping in the classical limit. The classical limit of
(\ref{Lrhog}) is not correct!

To try to understand better the origin of the difficulties with the purported equation of motion (\ref{Lrhog}), let us
consider how it might be ``derived''. Notice that (\ref{Lrhog}) would result from the undamped evolution equation
(\ref{evWig}) by the substitution \ben \rhog\ \meq\ T\left(\,\rho\,\right)\ ,\label{rhogTrho}\een a generalization of
(\ref{rgEnTrEn}). This simple identification is appealing in part because it is similar to (\ref{HusimiWT}), that defining the Husimi
phase-space distribution $\rho\sub H(p,q;t)$. The Husimi equation of motion is found directly from that for the Wigner function, by substituting (\ref{HusimiWT}). Additional terms arise in the Husimi equation of
motion compared to that for the Wigner function, since the coarse-graining evolves in time \cite{T}. Just as the ``twisting'' by $T_H$ modifies the equation of motion, so does application of the Dito-Turrubiates
transition operator $T$.  The
physical damping is meant to be introduced that way.

Following (\ref{smoothHW}), one might hope for an interpretation of $T(\rho)$ as the
Wigner distribution coarse-grained in momentum space. That point of view is not sensible, however, since the weighting would have to be Gaussian-like with an imaginary exponent. This does point to the origin of the main problem, however: unlike the Husimi
transition operator $T_H$, the Dito-Turrubiates $T$ is not real.

As a consequence, eqn. (\ref{TKE}) can hold: $T$  transforms a real Hamiltonian into a complex one. That is the crucial point here, we believe.  Compare to a
similar situation -- if a non-self-adjoint Hamiltonian is used, the equation of motion of the density operator is
modified,\footnote{\, See \cite{Es}, for example, where the analogous modified equation of motion for a Heisenberg
operator was shown to lead to a quantum anomaly.} to \ben 0\ =\ i\hbar\, \frac{\partial\hat\rho}{\partial t}\ +\
\hat\rho\, \hat H\ -\ \hat H^\dagger\, \hat\rho\ . \label{evhrHd}\een By analogy, we should consider\,\footnote{\, This
form may possibly be related to the bi-orthogonal quantum mechanics discussed by Curtright and Mezincescu \cite{CM}.
For related work in phase space, see \cite{CV} and \cite{SG}.}
 \ben  -\ i\hbar\,
\frac{\partial\rhog}{\partial t}\, \ =\ \rhog\, \stg\, H\ -\ \overline{\rhog\, \stg\, H}\  .\label{sarhoev}\een Notice
that the right-hand side of this equation is purely imaginary, so that \ben \frac{\partial\rhog}{\partial t}\, \ =\ -\ \frac 2
\hbar\, {\rm Im}\left(\, \rhog\, \stg\, H\ \,\right)\ .\label{sarhoevIm}\een This implies that the reality of the Wigner function \ben   \rhog\ =\ \overline{\rhog}\
\label{rhogreal}\een is preserved in evolution.\footnote{\, According to (\ref{sarhoevIm}), if $\rhog$ had a non-zero imaginary part, it would not evolve.} We can therefore also write \ben  -\ i\hbar\, \frac{\partial\rhog}{\partial t}\, \ =\
\rhog\, \stg\, H\ -\ H\, *_{-\gamma}\, \rhog\ .\label{sarhoevnobar}\een  With this prescription it is easy to show that
the classical limit makes sense for the simple harmonic oscillator Hamiltonian: \ben \lim_{\hbar\to 0}\, \frac{\rhog\,
\stg\, H\, -\, H\, *_{-\gamma}\, \rhog}{i\hbar}\, =\, \frac{ \{\rhog,H\}_\gamma\, -\, \{H,\rhog\}_{-\gamma} }2\, =\, \{
\rhog,H \}_\gamma\ .\label{climevr}\een We emphasize that the relation (\ref{rhogTrho}) is then {\it not} obeyed.

Let us now attempt to argue for (\ref{sarhoevnobar}) from other grounds. As a starting point, let us assume that the
Liouville Theorem still holds, and try to modify the argument that led to (\ref{evWig}) to justify the damped equation
of motion (\ref{sarhoevnobar}). An important advantage of the Dito-Turrubiates method is  that it only modifies the
classical brackets. That advantage would be lost if we need to input something to replace the Liouville Theorem.
Therefore, we assume the equation of motion is \ben \frac{d\rho_{\gamma,c}}{dt}\ =\ \frac{\partial
\rho_{\gamma,c}}{\partial t}\ +\ \{\, \rho_{\gamma,c}, H \,\}_\gamma\ =\ 0\ . \label{Liouvrhcd}\een The phase-space
version of the Dirac quantization rule (\ref{Dirps}) above should therefore be deformed to  \ben \{a, b\}_\gamma\ \rightarrow\
\frac{a\,\stg\,b\, -\, \overline{a\,\stg\,b}}{i\hbar}\ =\ \frac{a\,\stg\,b\, -\, b\,*_{-\gamma}\,a}{i\hbar}\
\label{Dirpsg}\een for real observables $a$ and $b$, in order to recover the equation of motion (\ref{sarhoev}).

One might be tempted to write \ben \rhog(p,q;t)\ \meq\ U\sub\gamma(p,q;t)\,\stg\,\rhog(p,q;0)\,
*\sub{-\gamma}\,\overline{U\sub\gamma(p,q;t)}\ \label{DTUgrgbUg}\een as a real solution to the equation of motion
(\ref{sarhoevnobar}). But this expression is ambiguous at best, as shown by the discussion around eqn. (\ref{foneh}).

Luckily, however, a formal solution to (\ref{sarhoevnobar}) {\it can} be written, by deforming the undamped solution of  (\ref{defadst}-\ref{expadst}). If we define \ben {\rm ad}^{(\gamma)}\sub *[f]\,g\ :=\ f\,\stmg\,g\ -\ g\,\stg\,f\ , \label{defadstg}\een then (\ref{sarhoevnobar}) is \ben
i\hbar\, \frac{\partial \rhog}{\partial t}\ =\ {\rm ad}^{(\gamma)}\sub *[H]\,\rhog \ =:\ i\hbar{\cal L\sub\gamma}\, \rhog\  .
\label{evWigadg}\een The solution is just \ben \rhog(p,q;t)\ =\ \exp\left\{\,\frac{-it}{\hbar}\, {\rm ad}^{(\gamma)}\sub *[H]
\,\right\}\, \rhog(p,q;0)\ =\ \exp\left\{\,t\,{\cal L}\sub\gamma \,\right\}\, \rhog(p,q;0)\ . \label{expLg}\een
There is no associative ambiguity in this explicit solution.

A more  explicit result can be given immediately for the damped harmonic oscillator, having \ben {\cal L}_\gamma\ =\ m\omega^2q\,\frac\partial{\partial p}\ -\ \frac p m \left( \frac\partial{\partial q}\ -\
2m\gamma \frac\partial{\partial p} \right)\ . \label{LgaSHO}\een The quantum evolution is
(\ref{sarhoevnobar}), but that reduces to (\ref{Liouvrhcd}), with $\rho_{\gamma,c}\to \rhog$. The quantum Wigner function satisfies the classical equation
of motion. The Wigner function at time $t$ is therefore simply \ben \rho\sub\gamma(p,q;t)\ =\ f\left(\, p_c(-t),q_c(-t)
\,\right)\ , \label{cflow}\een where $p_c(t)$ and $q_c(t)$ characterize the classical (damped) trajectories in phase
space. Thus the quantum damped harmonic oscillator follows the classical backward flow
of the phase-space coordinates. Perhaps this is not surprising, since for any quadratic Hamiltonian, such as that of the undamped simple
harmonic oscillator, the same results holds (see \cite{ZFC}, e.g.).

\section{ Discussion }

We first summarize. We have shown how to describe the dynamics of Wigner functions in the Dito-Turrubiates scheme \cite{DT}.
The non-Hermitian damped star product $\stg$ is c-equivalent to the Moyal product $*$. Therefore, if the evolution equation of the damped Wigner function only involves $\stg$-commutators, only Poisson brackets survive in the classical limit, rather than the damped bracket $\{\cdot,\cdot\}_\gamma$. The classical limit would then be damping-free, and therefore incorrect.

However, the damped transition operator $T$ transforms the simple harmonic oscillator Hamiltonian into a complex one, according to (\ref{TKE}). Consequently, the evolution equation does not involve $\stg$-commutation, but must instead be (\ref{sarhoevnobar}). This ensures that a real Wigner function remains real as it evolves, and the classical
limit is correct. Not only is the classical limit correct, it is exact. The damped harmonic oscillator in this
formalism therefore follows the classical flow, a property shared with non-dissipative systems in phase space having quadratic Hamiltonians.

Most helpful to us were (i) comparisons of the damped star product $\stg$ with other physical star products that are also c-equivalent to the Moyal $*$, viz. the standard product $*\sub S$, and the Husimi product $*\sub H$; and (ii) the Heisenberg equation of motion for an operator observable modified for a non-Hermitian Hamiltonian (see \cite{Es}, e.g.).

Let us now conclude with  a few of the many questions that remain. We hope progress can be made toward their answers.

{\it Should the damped $\stg$-product be obtained by
exponentiating $i\hbar\{\cdot,\cdot\}_\gamma/2$?} In the case at hand, much of the structure of the $\gamma$-deformed star product is irrelevant. Since we use the quadratic Hamiltonian (\ref{dhoH}),  the only terms that enter are those that are up to quadratic in $\lp$ and $\rp$. In this sense, our main result has a certain robustness. More work on this question would be helpful, however.

{\it Is there a deformed structure analogous to the Heisenberg-Weyl group that is relevant to the damped case?} In the undamped case, the Heisenberg-Weyl group is the structure of paramount importance. The Moyal $*$-product simply provides a $*$-realization of that group.  As discussed around  (\ref{foneh}), associativity can be a problem in the damped case. But if (\ref{expLg}) is an appropriate guide, perhaps \ben  \exp\left\{\, {\rm ad}^{(\gamma)}\sub *[ap+bq]    \,\right\}\, e^{cp+dq}\ =\  e^{cp+dq}\, e^{-i\hbar(ad-bc +2m\gamma ac)}\ \label{HWg}\een can serve the purpose. Here $a,b,c$ and $d$ are constants, so that $e^{ap+bq}$ and $e^{cp+dq}$ $*$-represent elements of the Heisenberg-Weyl group. When $\gamma\to 0$ in (\ref{HWg}), a form of the defining $*$-relations of the Heisenberg-Weyl group is recovered.

{\it Can a solution to the damped equation of motion (\ref{sarhoevnobar}) analogous to the undamped formula (\ref{Wig0t}) be
written?} It is clear from (\ref{expLg}) that the $*$-eigen equation of ${\rm ad}^{(\gamma)}\sub *[H]$ is relevant, so one can start there.  One possible ansatz follows, although it may only have relevance for very small $\gamma$. Suppose
we find the time-independent off-diagonal Wigner matrix elements $\rho\sub{\ga;\cE,\cE'} = \rho\sub{\ga;\cE,\cE'}(p,q)$
satisfying \ben H\,\stmg\,\rho\sub{\ga;\cE,\cE'}\ =\ \overline{\cE}\, \rho\sub{\ga;\cE,\cE'}\ ,\ \ \
\rho\sub{\ga;\cE,\cE'}\, \stg\, H\ =\ \cE'\,\rho\sub{\ga;\cE,\cE'}\ ,\label{rhocEEp}\een where the complex
eigenvalues\footnote{\, For the role such complex eigenvalues play in another description of the damped harmonic
oscillator, see \cite{Cii}. Wigner functions involving such eigenvalues are considered in \cite{Ci}.} are\ben \cE\ =\
E+i\lambda\ ,\ \ \cE'\ =\ E'+i\lambda'\ , \label{cEEla}\een with $E,E'\in \R$, and $\lambda,\lambda'\in\R_+$. If the real
eigenfunctions of (\ref{rhocEEp}) exist and are complete at all times $t$, then we can expand \ben \rhog(p,q;t)\ =\
\sum_{\cE,\cE'}\, R\sub{\cE,\cE'}(t)\, \rho\sub{\ga;\cE,\cE'}(p,q)\ . \label{Wig0tdexp}\een The equation of motion then
yields
 \bea \rhog(p,q;t)\
=\ \sum_{\cE,\cE'}\, R\sub{\cE,\cE'}(0)\, \exp\left[\frac{-i(\overline{\cE}-\cE')t}{\hbar} \right]\,
\rho\sub{\ga;\cE,\cE'}(p,q)\ \qq\nn\quad =\ \sum_{\cE,\cE'}\, R\sub{\cE,\cE'}(0)\, \exp\left[\frac{-i(E-E')t}{\hbar}
\right]\,\exp\left[\frac{-(\lambda+\lambda')t}{\hbar} \right]\, \rho\sub{\ga;\cE,\cE'}(p,q)\ . \label{Wig0td}\eea  If
this conjecture is correct, then the physically important $*$-eigen equations are those given in (\ref{rhocEEp}). It
would not be clear then that the phase-space functions of (\ref{rgEnTrEn}) are directly relevant.

The ansatz of (\ref{rhocEEp}-\ref{Wig0td}) reduces to the undamped system when $\hbar\to 0$. It is also devoid of
unphysical stationary states -- the imaginary parts of eigenvalues $\lambda$ and $\lambda'$ {\it both} contribute to
the damping of the dynamics. Furthermore, the system is consistent with (\ref{cflow}) if $(\overline{\cE}-\cE')/\hbar$
is independent of $\hbar$. Such eigenvalues were obtained in \cite{DT}, as indicated in (\ref{Egn}). However, the
corresponding solutions (\ref{rgEnTrEn}) are not real -- see (\ref{rgEdhoz}), e.g.

{\it Can the Dito-Turrubiates scheme be related to other quantization methods for dissipative systems (see \cite{D,Raz,W})?} Certainly, hints of connections with other models are apparent. For example, Dekker's use of a complex Hamiltonian is recalled by $T(H) = H -i\hbar\gamma/2$. Bateman's doubled system of a damped and anti-damped oscillator comes to mind from ${\rm ad}^{(\gamma)}\sub *[H]\rho = H\stg\rho-\rho*_{-\gamma}H$; in this last expression, however, the anti-damped ($\gamma\to -\gamma$) system is dual rather than extra/auxiliary. The importance of resonances has been emphasized by \cite{Ci,Cii} and others, and seems relevant to the $\stg$- and $\stmg$-eigen equations of the previous paragraph. Finally, if the correspondence (\ref{rhogTrho}) were correct, then we would be able to define a damped Wigner transform
\ben \cW\sub\gamma\ :=\ T^{-1}\,\cW\ =\ \cW\,\hat T^{-1}\ ,  \label{ThatdWig}\een  where \ben   \hat T^{-1}\ =\ \exp\left\{\,  -\frac{im}{2\hbar}\,\Big( {\rm ad}[\hat x]\Big)^2 \,\right\} . \label{hatT}\een The form of this $\hat T$ appears to be consistent with Tarasov's \cite{Tar} proposal to include superoperators in the operator formulation of quantum mechanics in order to adapt it to dissipative systems.

{\it Can other dissipative physical systems be treated in a similar way?} One very useful example might be spin systems, with relaxation.

{\it Instead of dissipation, might other physical effects, such as decoherence, be describable in the Dito-Turrubiates manner?}

%\vskip1cm\appendix

\vskip1cm\noindent{\bf Acknowledgements}\hfill\break We thank our colleagues Saurya Das, Arundhati Dasgupta, and Sourav
Sur for useful discussion. M.A.W. also thanks M. Razavy for a helpful conversation. This work was supported in part by
a Discovery Grant from the Natural Sciences and Engineering Research Council (NSERC) of Canada.

%%%%%%%%%%%%%%%%%%%%%%%%%%%%%%%%%%%%%%%%%%%%%%%%%%%%%%%%%%%%%%%%%%%%%%%%%%%%%%%%%%%%%%%%%%%%

\end{document}